\documentclass{amsart}
\usepackage{amssymb}
\usepackage{amsmath}
\usepackage{amsfonts}
\usepackage{graphicx}
\usepackage{epstopdf}

\usepackage{ytableau}
\usepackage[vcentermath]{youngtab}

\usepackage{float}
\usepackage{placeins}

\newcommand{\beq}{\begin{equation}}
\newcommand{\eeq}{\end{equation}}

\begin{document}

\begin{center}
{\Large\bf On the universal Casimir spectrum}\\
\vspace*{1 cm}
{\large  }
\vspace*{0.5 cm}

{\large  R.L.Mkrtchyan 
}

\vspace*{0.5 cm}

{\small\it Alikhanyan National Science Laboratory (Yerevan Physics Institute), \\ 2 Alikhanian Br. Str., 0036 Yerevan, Armenia}

{\small\it E-mail: mrl55@list.ru}

\vspace*{0.5 cm}

\end{center}

 {\bf Abstract.}  
 We conjecture the connection between $su$ and $so$ members of universal, in Vogel's sense, multiplets. 
 The key element is the  notion of the {\it vertical componentwise sum}  $\oplus_v$ of Young diagrams. Representations in the decomposition of the power of the adjoint representation of $su(N)$ algebra can be parameterized by a couple of $N$-independent Young diagrams $\lambda$ and $\tau$, with equal area.  We assume that the $so(N)$  member of the universal (Casimir) multiplet of a given $su(N)$ representation is the $so$ representation with $\lambda \oplus_v \tau$ Young diagram. This  allows one to obtain the universal form of the Casimir eigenvalue on that multiplet.  
Conjecture  is checked for all known cases: universal decompositions of powers of adjoint up to fourth,  and series of universal representations.   

On this basis we suggest the set of universal Casimirs for fifth power of adjoint. We also conjecture that vertical sum operation is a kind of the (dual version of the) folding map of Dynkin diagrams. This will hopefully explain the intrinsic symmetry of universal formulae with respect to the automorphisms of Dynkin diagrams.

\vspace*{1 cm}

\section{Universal multiplets}

Consider Vogel's \cite{V0} universal formula for dimensions of the adjoint representations of simple Lie algebras $\mathfrak{g}$, and universal eigenvalue of Casimir operator on them, in terms of homogeneous parameters  $\alpha, \beta, \gamma$ on projective plane:

\begin{eqnarray} \label{ad}
dim \, \mathfrak{g} &=&	\frac{(\alpha-2t)(\beta-2t)(\gamma-2t)}{\alpha \beta \gamma} \\ \nonumber
	C&=&2t
\end{eqnarray}

This dimension formula on the points from Vogel's table (see Appendix,  table \ref{tab:Vogel}) gives dimensions of adjoint representations of corresponding algebra. So these representations become connected not only by the common name and meaning, but also by this universal formula. 

Vogel's theory is invariant w.r.t. the permutation (and rescaling) of universal parameters. 
Since expression  above is symmetric w.r.t. the permutation of  parameters, from each algebra in this correspondence participates one, adjoint, representation.  Similarly expression for Casimir gives corresponding eigenvalues for each algebra, exactly one, since it also is symmetric. 

Consider universal expressions for dimensions and eigenvalues for representations in the tensor square of adjoint. 
The symmetric square  decomposes universally \cite{V0} as:

\begin{eqnarray}\label{sad}
	S^2 \mathfrak{g}=1 \oplus Y_2(\alpha) \oplus Y_2(\beta) \oplus Y_2(\gamma)
\end{eqnarray}

Dimensions of representations and their Casimirs are given by the universal formulae

\begin{eqnarray} \label{dsad}
	\text{dim} \, Y_2(\alpha)&=& \frac{\left(  2t  - 3\alpha \right) \,\left( \beta - 2t \right) \,\left( \gamma - 2t \right) \,t\,\left( \beta + t \right) \,
		\left( \gamma + t \right) }{\alpha^2\,\left( \alpha - \beta \right) \,\beta\,\left( \alpha - \gamma \right) \,\gamma} \\ \nonumber
		C&=&4t-2\alpha
\end{eqnarray}
and their transformations under permutations of  parameters. 

Similarly, for the antisymmetric square one has decomposition  \cite{V0}:

\begin{eqnarray}\label{daad}
	\wedge^2 \mathfrak{g}&=&\mathfrak{g} \oplus X_2 \\ \nonumber 
	\text{dim} X_2&=&\frac{(2t-\alpha)(2t-\beta)(2t-\gamma)(t+\alpha)(t+\beta)(t+\gamma)}{\alpha^2 \beta^2 \gamma^2} \\  \nonumber
	C&=&4t
\end{eqnarray}

We see that universal formulae (\ref{dsad}), (\ref{daad}) combine together different representations of a given algebra, provided we specialize universal parameters for that algebra (and their permutations), and also combine representations of different algebras, provided we take parameters for different algebras. For a specific universal dimension formula, we shall call the set of representations of an algebra, appearing at different permutations of parameters, the {\it small (universal) multiplets of a given algebra}, and the set of  small multiplets for all algebras we shall call  {\it big (universal) multiplets}. E.g. there is small multiplet for each algebra consisting from one adjoint representation, and big multiplet of these small representations is the set of all adjoint representations. Similarly, considering the symmetric square of adjoint, we find small multiplets consisting  from the three representations for classical algebras, and two representations for exceptional algebras, see table \ref{tab:ad2} (where we take as a representative of exceptional algebras the algebra $E_7$), and altogether they form the big multiplet. 

In the table \ref{tab:ad2} and below we use notation $D(\lambda,\tau)$ for $su(N)$  representations with the following Dynkin labels    
\begin{eqnarray}
D(\lambda,\tau)=(\lambda_1, \lambda_2, ...,\lambda_k, 0,...,0, \tau_k,...,\tau_2,\tau_1)
\end{eqnarray}
with fixed (i.e. independent of $N$) $k$. We actually need symmetrized w.r.t. the $Z_2$ automorphisms of Dynkin diagram of $su$ representation $D_s(\lambda,\tau)=D(\lambda,\tau)\oplus D(\tau,\lambda)$ for $\lambda \neq \tau$, and $D_s(\lambda,\lambda)=D(\lambda,\lambda)$. Sometimes it is convenient to use notation  $(\lambda_1, \lambda_2, ...,\lambda_k, 0,...,0, \tau_k,...,\tau_2,\tau_1)_s=D_s(\lambda,\tau)$. We also  identify in a standard way the set of  labels $(\lambda_1, \lambda_2,...)$  with Young diagram $\lambda$, and in the same way the set of labels $(\tau_1, \tau_2,...)$ with Young diagram $\tau$. Also everywhere we assume $N$ to be "large enough" for antisymmetric tensors to be non-zero.

 $Y_2(.)$ representations of table \ref{tab:ad2} form a big multiplet of representations in the symmetric square of adjoint, and there is no other multiplets in that square, excluding trivial singlet one (i.e. one singlet for each algebra). Similarly in the antisymmetric square we find one big multiplet of adjoint representations, and one big multiplet of $X_2$ representations, which is similar to adjoint one, i.e. its universal formulae are symmetric under permutations of parameters.

\ytableausetup{smalltableaux}

\begin{table}
	\caption{Square of adjoint}
	\label{tab:ad2}
	
	\begin{tabular}{|c|c|c|c|c|}
		\hline
		Irrep	&  Casimir&su  &so  & $E_7$ \\ 
		\hline
		$\mathfrak{g}$	&  $2t$ &  $D_s\left( \ydiagram{1},\ydiagram{1} \right)$  &  \ydiagram{1,1} & {\bf 133} \\
		\hline
		$Y_2(\alpha)$	& $4t-2\alpha$ & $D_s\left(  \ydiagram{2},\ydiagram{2} \right) $   & \ydiagram{2,2} & {\bf 7371 }\\
		\hline
		$Y_2(\beta)$	& $4t-2\beta$ &$D_s \left(   \ydiagram{1,1},\ydiagram{1,1} \right) $ & \ydiagram{1,1,1,1}  & {\bf 1539 }\\
		\hline
		$Y_2(\gamma)$	& $4t-2\gamma$ & $ D_s \left( \ydiagram{1},\ydiagram{1}\right)$ &  \ydiagram{2}&  0 \\
		\hline
		$X_2$	& $4t$ &$D_s\left( \ydiagram{2},\ydiagram{1,1} \right)$& \ydiagram{2,1,1} & {\bf 8645} \\
		\hline
	\end{tabular}
\end{table}


Each universal dimension formula gives one big multiplet, and corresponding small multiplets.  Note also that the same representation of a given algebra can be the member of a different small (and hence different big) multiplets, as we see in the table \ref{tab:ad2} for adjoint representation of $su$. It is the member of  multiplet of adjoint representations, according  to (\ref{ad}), and also is the member $Y_2(\gamma)$ of the $su$  small multiplet of symmetric square. Actually, in Vogel's picture \cite{V0}, it was assumed that all these representations can be defined (as some categories) at an arbitrary values of parameters, and they are generally non-zero and different. But  for a given simple Lie algebra some of them can coincide, as we just observed.

For each  big multiplet we have also a universal Casimir expression, which gives the values of Casimir operator on each representation of big multiplet by specializing on the point of Vogel's table, and by permutation of universal parameters. However, note that the value of Casimir operator on zero representations is not fixed, and can be defined arbitrarily, even different for "different zero representations" (i.e. appearing in different universal multiplets). And indeed we see that e.g. for zero representation of exceptional small multiplet of $E_7$ in table \ref{tab:ad2} the value of Casimir operator, $4t-2\gamma=60$, is non-zero and seemingly not having the meaning purely within algebra. To illustrate the mechanism, consider  e.g.  the trace of power of Casimir operator on some representation. It is equal to 

\begin{eqnarray}\label{dim0}
	Tr C^n= \sum_{i} c_i ^n dim_i
\end{eqnarray}
where $i$ enumerates eigenspaces of $C$ on that representation, $c_i$ are corresponding eigenvalues of $C$, and $dim_i$ are dimensions of corresponding eigenspaces. As it is clear from (\ref{dim0}), if for some $i$ $dim_i=0$ then the value of $c_i$ becomes irrelevant. 

There is another  type of universal multiplets, which we shall call {\it Casimir small (universal) multiplets (of a given algebra)}, and respectively  {\it Casimir big (universal) multiplets}. As an example consider  the  $X_3$ representation in the Vogel's universal decomposition of cube of adjoint representation.  It appears in the antisymmetric subspace and is irreducible for exceptional algebras and reducible for classical ones, see table \ref{tab:X3}, and \cite{V0,Cohen}. 

\ytableausetup{smalltableaux}

\begin{table}
	\caption{$X_3$}
	\label{tab:X3}

\begin{tabular}{|c|c|c|c|c|}
	\hline
	Representation &  C & $su$ & $so$ & $E_7$ \\
	\hline
$X_3$	& $6t$ & $D_s(\ydiagram{3},\ydiagram{1,1,1}) \oplus  D_s(\ydiagram{2,1},\ydiagram{2,1}) $ & $\ydiagram{3,1,1,1} \oplus \ydiagram{2,2,2}$  & 365750 \\
	\hline
\end{tabular}
\end{table}


This representation has a universal dimension formula:

\begin{eqnarray}\label{x3dim}
	\frac{1}{6} dim \, \mathfrak{g} (dim \, \mathfrak{g}-1)(dim \, \mathfrak{g}-8)
\end{eqnarray}

For classical algebras it is reducible, however, to obtain the universal expressions for its irreducible subspaces, one need an extension of the ring of coefficients  \cite{V0}. Moreover, the possibility of decomposition of further Casimir subspaces is doubtful, due to the properties of that ring of coefficients,  since it has a zero divisors and even the complete reducibility is questionable. 

For these reasons,  in \cite{AIKM} it is suggested to study the question of universality on the level of Casimir subspaces, i.e. subspaces with the same Casimir eigenvalue. In that case there is no need to decompose $X_3$ into irreducible subspaces, since it itself is already Casimir subspace, and it has a universal dimension formula, (\ref{x3dim}). 

Now the definition of Casimir small  multiplets is clear, they are sets of irreps with the same universal Casimir's values, connected by permutation of universal parameters within one algebra. Respectively the union of these sets  for all algebras constitute the big Casimir (universal) multiplet. 

\section{The problem}

Our aim is to find out weather all eigenvalues of Casimir on an arbitrary power $\mathfrak{g}^k$ of adjoint can be represented in the universal form, as it is given in e.g. table \ref{tab:ad2} for the square $\mathfrak{g}^2$.  Of course that list includes universal Casimirs already appeared for the powers $<k$, and new ones, specific for power $k$. After some considerations in this and subsequent sections, we suggest some conjectures in section \ref{Conj}. 

The general form of universal Casimir

\begin{eqnarray}\label{UC}
	C=x\alpha+y\beta+z\gamma 
\end{eqnarray}

contains three unknown coefficients $x,y$ and $z$.  Assume that we know the $su$ and   $so$ members of universal multiplet and correspondingly (\ref{UC}) gives the values of their Casimirs, when specialized on corresponding algebra. For $so(N)$ the general form of Casimir's eigenvalue of some representation $\Lambda$  is 
$AN+o$, where $A$ is the area of Young diagram of $\Lambda$, $o$ is some integer.  Substituting the values of universal parameters for $so(N)$ we get an equations 

\begin{eqnarray}
	z=A, \\
	-2x+4y-4z=o
\end{eqnarray}

For $su(N)$ representations $D_s(\lambda,\tau)$ the value of corresponding Casimir is 

\begin{eqnarray}
	(A_\lambda+A_\tau)N +u+ \frac{1}{N} (A_\lambda-A_\tau)^2
\end{eqnarray}
where $A_\lambda$ and $A_\tau$ are areas of $\lambda$ and $\tau$ diagrams, respectively, $u$ is some integer. 
As is noticed in \cite{MRL25}, this expression can coincide with universal Casimir, which for $su(N)$ algebra is equal to
$-2x+2y+zN$, iff  $A_\lambda=A_\tau $, which is exactly  the case for irreps appearing in the powers of adjoint of $su(N)$. Parameters in that case are determined to be 

\begin{eqnarray}
	A_\lambda=A_\tau \equiv k \\
	z=2k \\
	-2x+2y=u
\end{eqnarray}

Altogether we obtain conditions for universal Casimir to give the values of Casimirs of two representations:  $\Lambda$ of $so(N)$ and $D_s(\lambda,\tau)$  of $su(N)$. Namely, the areas of diagrams should be equal: $A_\lambda=A_\tau\equiv k$, then $A$ should be equal to $2k$, as is the case, and parameters become equal to

 \begin{eqnarray}\label{UCus}
 		C&=&x\alpha+y\beta+z\gamma \\
 	x&=&2k+\frac{1}{2}o-u, \\
 	y&=&2k+\frac{1}{2}(o-u), \\
 	z&=&2k
 \end{eqnarray}

So, for $k$-th power of adjoint $\mathfrak{g}^k$, one can determine universal Casimir by knowing two members of big universal multiplet, $su$ and $so$. 

In the next section we shall present recipe, how one can find $so$ member of universal big multiplet, corresponding to given $su$ representation, and check it on a number of cases (actually all known). So, one can decompose power of adjoint of $su$, and then derive a large number of universal Casimirs.

\section{The vertical  sum of Young diagrams}

We first define  the {\it vertical componentwise sum  $\oplus_v$ of Young diagrams}. 

Put diagrams $\lambda$ and $\tau$  one under the other, left vertical lines  vertically aligned, and push up any box from lower diagram until it meet some box higher or the upper horizontal line of upper diagram. Result is another Young diagram, with number of boxes equal to the sum of the numbers of boxes in the initial diagrams. 

Another description: denote $h^\mu_i$ the height of the $i$-th column of Young diagram $\mu$ (numbered, e.g. from left to the right). These heights completely characterize the diagram. Then $h^{\lambda \oplus_v \tau}_i=h^\lambda_i+h^\tau_i$. 

Third description: denote $\{ l_i \}$  the (unordered) set of lengths of all rows in diagram  $\lambda$, this set also completely characterize the diagram. Let $\{t_\alpha\}$ be the same for diagram $\tau$, then $\lambda \oplus_v \tau$ has rows with the set of lengths  $\{l_i, t_\alpha\}$.

We denote this map of couple of representations $\lambda, \tau$ to $\lambda\oplus_v\tau$ by $M$:

\begin{eqnarray}
	M: (\lambda,\tau) \rightarrow \lambda\oplus_v\tau
\end{eqnarray}

Pictorially, for some "general" diagrams

\begin{eqnarray}
	\lambda &=& \ydiagram{3,3,1} \\
	\tau&=&\ydiagram{4,2,1} 
\end{eqnarray}
	we have 
	\begin{eqnarray}
	\lambda\oplus_v\tau&=&\ydiagram{4,3,3,2,1,1}
\end{eqnarray}

Obviously, the vertical sum is commutative: 

\begin{eqnarray}
	\lambda\oplus_v\tau = \tau\oplus_v\lambda
\end{eqnarray}

If we introduce $k$-th Cartan power $k\lambda$ of a given representation with Young diagram $\lambda$, its diagram will be that of $\lambda$ with  $k$ times more columns of given height. It is easy to show that 
\begin{eqnarray}
	k\lambda\oplus_v k\tau = k(\lambda\oplus_v \tau)
\end{eqnarray}

Actually we notice some resemblance of this operation with so-called folding \cite{Ster,KacInf} of algebras whose   Dynkin diagrams have a automorphisms symmetry - $A_n, D_n, E_6$, but we shall not study that observation in the present paper. 

Few examples with small number of boxes.

Example 1:
\begin{eqnarray}
	\lambda &=& \ydiagram{1}  , \,\,\,
	\tau=\ydiagram{1} \\
	\lambda\oplus_v\tau&=&\ydiagram{1,1}
\end{eqnarray}
i.e. from the adjoint of $su$ we obtain the adjoint of $so$. 

Example 2:
\begin{eqnarray}
	\lambda &=& \ydiagram{2}  , \,\,\,
	\tau=\ydiagram{2} \\
	\lambda\oplus_v\tau&=&\ydiagram{2,2}
\end{eqnarray}
i.e. the Cartan square of $su$ adjoint gives the  Cartan square of adjoint of $so$.

Example 3:
\begin{eqnarray}
	\lambda &=& \ydiagram{1,1}  , \,\,\,
	\tau=\ydiagram{1,1} \\
	\lambda\oplus_v\tau&=&\ydiagram{1,1,1,1}
\end{eqnarray}

\section{Construction of universal Casimirs}

One start from an arbitrary representation $D_s(\lambda,\tau)$ with equal area $\lambda$ and $\tau$, construct the  diagram $\Lambda=\lambda\oplus_v\tau$, take corresponding irrep of corresponding $so$ representation (which we also denote $\Lambda$), calculate  Casimirs of $D_s(\lambda,\tau)$ and $\Lambda$, and derive a universal Casimir by formulae  (\ref{UCus}). 

Consider this construction for Examples 1,2,3 above, and for some infinite series of $su$ diagrams.

{\bf Example 1. }

Casimir of adjoint of $su(N)$ is $2N$, for $so(N)$ it is $2N-4$, so 

\begin{eqnarray}
	C=2\alpha+2\beta+2\gamma 
\end{eqnarray}
i.e. we obtain well-known universal expression for Casimir  eigenvalue of adjoint representations, which means that universal Casimir eigenvalues give correct answer for all algebras. 

{\bf Example 2.} 

Casimir of $su$ is 4N+4, of $so$ it is $4N-4$, so 

\begin{eqnarray}
	C=2\alpha+4\beta+4\gamma 
\end{eqnarray}

This Casimir coincides, up to permutations, with that of next Example 3. 

{\bf Example 3.} 

Casimir of $su$ is $4N-4$, of $so$ it is $4N-16$, so 

\begin{eqnarray}
	C=4\alpha+2\beta+4\gamma 
\end{eqnarray}

 Last two examples  coincide with known universal Casimir values for $Y_2(\alpha)$ and $Y_2(\beta)$, hence give correct answers for all algebras. 
 
 But there is third member of this small multiplet, namely $Y_2(\gamma)$, with Casimir $C=4\alpha+4\beta+2\gamma$. 
 For $su$ that is representation, which coincides with adjoint one, and indeed we obtain $C=2N$, as it should. For $so$ it is diagram
 
 \begin{eqnarray}\label{2}
 	 \ydiagram{2} 
 \end{eqnarray}

which  cannot be represented as a vertical sum of two diagrams with equal area. So, in principle, it is not evident that its Casimir eigenvalue will appear among the spectra of universal Casimirs. However, it is the case, since for $so$ we have  $C=4\alpha+4\beta+2\gamma=2N$, which is correct value for the representation (\ref{2}). 

Altogether, this universal Casimir also covers the eigenvalues of $Y_2(.)$ for exceptional algebras, too, as we know from Vogel's work \cite{V0}. Actually, as mentioned earlier, the value of universal Casimir on $Y_2(\gamma)$ for exceptional algebras is irrelevant, since dimensions of this representation is zero.

{\bf The $n$-th Cartan power of adjoint.}

The $n$-th Cartan power of adjoint of $su$ is $D(\sigma_n,\sigma_n)$, where $\sigma_n$ is the row of $n$ boxes. Its Casimir is 

\begin{eqnarray}
	C=2n^2+2nN-2n
\end{eqnarray}

The corresponding $so$ diagram is:

\begin{eqnarray}
	\sigma_n\oplus_v \sigma_n=
	\begin{ytableau}
		~& 	~ &\none[\cdot\cdot] &~ \\
		~&~&\none[\cdot\cdot]& 
	\end{ytableau}
\end{eqnarray}
where in the first and second rows there are $n$ boxes. This is the $n$-th Cartan power of adjoint. Corresponding Casimir of $so$ is 

\begin{eqnarray}
	C=2n^2+2nN-6n
\end{eqnarray}

so universal one is

\begin{eqnarray}
	C=2n\gamma+2n\beta-n(n-3)\alpha=\\
	2nt-n(n-1)\alpha
\end{eqnarray}

This coincides with universal Casimir given in \cite{LM1}, which gives correct answers for exceptional algebras, too.

The content of multiplets is revealed by applying permutation of parameters to universal Casimir. For large $k$ (here "large" is different for different algebras) Casimir becomes negative, and corresponding dimension formulae give zero, in agreement to that. For "small" $k$ one can find Casimir and find the corresponding representation, taking into account  the universal dimension formulae \cite{LM1}. 

{\bf Cartan product of $n$ adjoint and $k$ $X_2$}

Cartan product of $n$ adjoint and $k$ $X_2$ irreps of $su$ has Dynkin labels $(n+2k,0,...0,k,n)$ and Casimir eigenvalue $(4k+2n)N+6k^2+6kn-6k+2n^2-2n$. The corresponding vertical sum $so$ diagram has Dynkin labels $(k,n,k,0...)$, which are labels of $n$ adjoint and $k$ $X_2$ representations, but now for $so$ algebra,  and Casimir $(4k+2n)N+6k^2+6kn-14k+2n^2-6n$. From this, the universal Casimir is:

\begin{eqnarray}
	C=(4k+2n)\gamma+(4k+2n)\beta+(-3k^2+7k-3kn-n^2+3n)\alpha=\\
	(4k+2n)t+(-3k^2+3k-3kn-n^2+n)\alpha
\end{eqnarray}

in full agreement with previous calculation \cite{AM19} and particular case $k=0$ above.

{\bf $X_3$ representation.}

This is an example of  Casimir subspaces, such that its constituent irreps have no standard universal dimension formula (i.e. to have a dimension formula one needs an extension of the ring of universal parameters, see \cite{V0}).  In the cube of adjoint there is one such Casimir subspace, which consists of two irreps, for $su$ and $so$ algebras, and is irreducible for exceptional ones. 

For $su$ one of these representations has Dynkin labels $(3,0,...,0,1,0,0)_s$, and another has $(1,1,0,...0,1,1)$. Their Casimir is $6N$.

The corresponding $so$ representations are

\begin{eqnarray}
	\ydiagram{3,1,1,1} \,\,\,,\,\,\, \ydiagram{2,2,1,1}
\end{eqnarray}

Casimir of the first irrep is $6N-12$, of the second  $6N-16$

$X_3$ for $so$ consists from two representations, one is the first one of these diagrams, second is:

\begin{eqnarray}
	\ydiagram{2,2,2} 
\end{eqnarray}
with Casimir $6N-12$, so coincide with Casimir of the first diagram above, as it should, if they are members of the $X_3$.

We shall analyze $\mathfrak{g}^3$ case in details below, at this moment note that this last diagram cannot be obtained as a sum of two diagrams of equal area, perhaps that is a reason why it is in the $X_3$ of $so$. 

From the data above universal eigenvalue of Casimir on $X_3$ is $6t$, in agreement with Vogel \cite{V0}. Universal Casimir for second pair of irreps is $4t+2\gamma$ (it is B multiplet in notations of \cite{V0}).

\section{Powers of adjoint: 2, 3, 4}

\subsection{$\mathfrak{g}^2$}

Decomposition of $su(N)$  in terms of Dynkin labels:
\begin{eqnarray}
\mathfrak{g}^2=	\text{(0...0)}+\text{(010...02)}+\text{(010...010)}+  \\ 2 \text{(10...01)}+\text{(20...02)}+\text{(20...010)}
\end{eqnarray}

The same decompositions for $so(N)$ algebras:

\begin{eqnarray}
\mathfrak{g}^2=
	\text{(0...)}+\text{(010...)}+\text{(20...)}+ \\
	\text{(00010...)}+ 
	\text{(020...)}+\text{(1010...)}
\end{eqnarray}

Let's recover the universal form of these decompositions according to the conjecture. 

Start from area 4 su irreps. According to recipe, from e.g. (20...02) of $su$ we get (020...) of $so$. Corresponding Casimirs are $4N+4$, and $4N-4$, respectively. Universal Casimir then is $C=2\alpha+4\beta+4\gamma$. All this we already saw, this is the Cartan square of adjoint. Under permutations of parameters, this representations give rise to (010...010), and (10...01) for $su$, and to (00010...) and (20...) for $so$. 

Next we take the only remaining irrep of area 4 for $su$, namely (20...010), plus conjugate. It gives (1010...) for $so$. Casimir's are  $4N$ and 
$4N-8$, respectively. Universal Casimir is then $C=4\alpha+4\beta+4\gamma$. These are $X_2$ representations, for all algebras. 

Remaining irreps in both decompositions are one adjoint, and one singlet, which are universal. The obtained universal decompositions covers the case of exceptional algebras, too, as we know from Vogel \cite{V0}.

\subsection{$\mathfrak{g}^3$}

Decomposition of cube for $su$:
\begin{eqnarray*}
\mathfrak{g}^3=	2 \text{(0...0)}+\text{(0010...03)}+ \\  
2 \text{(0010...011)}+\text{(0010...0100)}+6 \text{(010...02)}+6 \text{(010...010)}+ \\ 9 \text{(100000001)}+2 \text{(110000003)}+4 \text{(110000011)}+2 \text{(110000100)}+  \\  6 \text{(200000002)}+6 \text{(20...010)}+\text{(30...03)}+  \\ 2 \text{(30...011)}+\text{(30...0100)}
\end{eqnarray*}

The same decomposition for $so$:

\begin{eqnarray*}
\mathfrak{g}^3=\text{(0...)}+\text{(0000010...)}+3 \text{(00010...)}+\\ 
\text{(0020...)}+6 \text{(010...)}+ 3\text{(01010...)}+ \\ 
3 \text{(020...)}+\text{(030...)}+2 \text{(100010...)}+ \\ 
6 \text{(1010...)}+  2 \text{(1110...)}+ \\
3 \text{(20...)}+\text{(20010...)}+3 \text{(210...)} 
\end{eqnarray*}

We carry on the  same procedure for this case and  combine data in the table \ref{tab:ad3}.  First column lists all $su$ representations, everywhere where needed it is implied symmetrized representation, i.e. $(....)_s$. Next column presents  their Casimirs, next are corresponding $so$ representations. In this column we also list remaining $so$ representations which have no preimage. In the next column we show corresponding Casimirs of $so$ representations,  then corresponding universal Casimir. Last column contains Vogel's notation for corresponding universal representation. Column of universal Casimirs  completely coincides with  Vogel's data. 

\begin{table}
		
\caption{Universal  Casimirs of $\mathfrak{g^3}$}
\label{tab:ad3}
\begin{tabular}{|c|c|c|c|c|c|} 
    \hline
	su & C  & so  & C  & C univ &  Vogel  \\
	\hline
	(30...0100)& 6N &(20010...)  & 6N-12 & $6\gamma+6\beta+6\alpha$ & $X_3$  \\
	\hline
	(30...011)&6N+6  &(1110...)  & 6N-6 & $6\gamma+6\beta+3\alpha$ & $C(\alpha)$  \\
	\hline
	(30...03)& 6N+12 &(030...)  &6N  &$6\gamma+6\beta$  &$Y_3(\alpha)$   \\
	\hline
	(110...100)& 6N-6 &(100010...)  &  6N-24& $6\gamma+3\beta+6\alpha$ &  $C(\beta)$ \\
	\hline
	(110...011)&  6N&(01010...)  &6N-16  & $6\gamma+4\beta+4\alpha$ & $B(\gamma)$  \\
	\hline
	(0010...0100)&  6N-12& (0000010...) &6N-36  & $6\gamma+6\alpha$ &$Y_3(\beta)$    \\
	\hline
	&  & (0020...) & 6N-12 & $6\gamma+6\beta+6\alpha$ & $X_3$  \\
	\hline
	&  & (210...) & 4N  &  $4\gamma+6\beta+4\alpha$ & $B(\beta)$  \\
	\hline
	(20...010)& 4N & (1010...) &4N-8  & $4\gamma+4\beta+4\alpha$ & $X_2$ \\
	\hline
	(20...02)&  4N+4&(020...)  & 4N-4 &$4\gamma+4\beta+2\alpha$  &$Y_2(\alpha)$  \\
	\hline
	(010...010)& 4N-4 &(00010...)  &  4N-16&  $4\gamma+2\beta+4\alpha$  & $Y_2(\beta)$  \\
	\hline
	(10...01)&  2N& (010...) & 2N-4 & $2\gamma+2\beta+2\alpha$   & $\mathfrak{g}$  \\
	\hline 
	& & (20...)& 2N & $2\gamma+4\beta+4\alpha$& $Y_2(\gamma)$\\
	\hline
\end{tabular}
\end{table}

We see from this table, that, as in the square of adjoint, there are diagrams of $so$ which have no preimage in $su$. But again, as in the square of adjoint, the Casimir of each such representation is covered by universal Casimir's appearing from other representations.

\subsection{$\mathfrak{g}^4$}

For the fourth power of adjoint one can carry on the same procedure, and compare with the results of \cite{AIKM}. Specifically, one recovers precisely the table 2 of \cite{AIKM} of all universal Casimirs. To make comparison one has to take into account that this is a table of eigenvalues of 4-split Casimir operator $C^{(4)}$. Connection with eigenvalues of our standard Casimir  $C$ is:

\begin{eqnarray}
	C=8t+4tC^{(4)}
\end{eqnarray}

For completeness, let us list diagrams, which cannot be represented as a vertical sum of appropriate $su$ diagrams, besides those which appear already at square and cube of adjoint:

\begin{eqnarray}
	\ydiagram{3,3,2}, \,\, \ydiagram{4,2}, \,\, \ydiagram{4}
\end{eqnarray}

Their Casimir's eigenvalues all appear from universal Casimirs, listed in \cite{AIKM}. 

\section{ Fifth power $\mathfrak{g}^5$}

In this Section we present the list of universal Casimirs, appearing in the fifth power of adjoint according to our recipe. First below we give the result of decomposition of fifth power of adjoint, in terms of Dynkin labels. This is calculated by LieART package of Mathematica\texttrademark.

\small
\begin{eqnarray*}
	44(000000000000)+265(100000000001)+320(010000000010)+320(200000000010)+\\
	320(010000000002)+320(200000000002)+130(001000000100)+130(300000000100)+\\
	130(001000000003)+260(110000000100)+260(001000000011)+130(300000000003)+\\
	260(300000000011)+260(110000000003)+20(000100001000)+520(110000000011)+\\
	(000010010000)+20(400000001000)+20(000100000004)+40(020000001000)+\\
	40(000100000020)+60(101000001000)+60(000100000101)+60(210000001000)+\\
	60(000100000012)+20(400000000004)+40(400000000020)+40(020000000004)+\\
	60(400000000101)+60(101000000004)+80(020000000020)+120(101000000020)+\\
	120(020000000101)+(500000010000)+(000010000005)+60(400000000012)+\\
	60(210000000004)+4(100100010000)+4(000010001001)+180(101000000101)+\\
	120(210000000020)+120(020000000012)+180(210000000101)+180(101000000012)+\\
	5(011000010000)+5(000010000110)+180(210000000012)+6(201000010000)+\\
	6(000010000102)+5(120000010000)+5(000010000021)+4(310000010000)+\\
	4(000010000013)+(500000000005)+4(500000001001)+4(100100000005)+\\
	16(100100001001)+5(500000000110)+5(011000000005)+20(100100000110)+\\
	20(011000001001)+4(500000000013)+4(310000000005)+5(500000000021)+\\
	5(120000000005)+6(500000000102)+6(201000000005)+16(310000001001)+\\
	16(100100000013)+24(201000001001)+24(100100000102)+20(120000001001)+\\
	20(100100000021)+25(011000000110)+20(310000000110)+20(011000000013)+\\
	25(120000000110)+25(011000000021)+30(201000000110)+30(011000000102)+\\
	16(310000000013)+20(310000000021)+20(120000000013)+24(310000000102)+\\
	24(201000000013)+25(120000000021)+30(201000000021)+30(120000000102)+\\
	36(201000000102)
\end{eqnarray*}
\normalsize

Actually, since we do not follow the multiplicities of representations, this decomposition can be described very simply: this is the sum with non-zero coefficients of representations $D_s(\lambda,\tau)$ with all possible $\lambda,\tau$ diagrams with equal area less or equal five. Perhaps, that is the case for higher powers, also. 

We calculate universal Casimirs originated from these representations according to our recipe and present them in the table \ref{tab:ad5}. Representations are identified by Dynkin labels, all representations are implied to be symmetrized, i.e. $(....)$ means $(...)_s$. Some of universal Casimirs in this table are permuted versions of other Casimirs from the table, there are nine such pairs, so total number of  independent (i.e. not connected by permutation of parameters) universal Casimirs in the table is nineteen. 
\FloatBarrier

\begin{table}
	\caption{Universal Casimirs in $\mathfrak{g}^5$} \label{Cas5}
	\label{tab:ad5}
	$\begin{array}{|c|c|c|c|}
		\hline
		\text{Repr}& \text{C univ} & \text{Repr} & \text{C univ} \\
		\hline
		(201000000102)&4 \alpha+4\beta+10\gamma & (120000000102) & 4\alpha+6\beta+10\gamma\\
		\hline
		(120000000021)& 2 \alpha+6\beta+10\gamma& (201000000013) & 2\alpha+7\beta+10\gamma \\
		\hline
		(120000000013)&   \alpha+8\beta+10\gamma &  	(310000000013)&   -2\alpha+8\beta+10\gamma \\
		\hline
		(011000000102)& 6 \alpha+4\beta+10\gamma  &(011000000021)  & 5 \alpha+5\beta+10\gamma  \\
		\hline
		(011000000013)&  4\alpha+7\beta+10\gamma  & (011000000110) & 6 \alpha+2\beta+10\gamma  \\
		\hline
		(100100000021)&  8\alpha+5\beta+10\gamma  & (100100000102) &  7 \alpha+2\beta+10\gamma \\
		\hline
		(100100000013)&  6\alpha+6\beta+10\gamma  &  (201000000005)& 10\beta+10\gamma   \\
		\hline
		(120000000005)& -2 \alpha+10\beta+10\gamma  & (310000000005) & -5\alpha+10\beta+10\gamma   \\
		\hline
		(011000001001)& 8 \alpha+\beta+10\gamma  & (011000000005) &  2\alpha+10\beta+10\gamma  \\
		\hline
		(100100001001)& 8 \alpha-2\beta+10\gamma  & (100100000005) &  5\alpha+10\beta+10\gamma  \\
		\hline
		(500000000005)& -10 \alpha+10\beta+10\gamma  & (000010000013) &10  \alpha+5\beta+10\gamma  \\
		\hline
		(000010000021)& 13\alpha+5\beta+10\gamma   & (000010000102) &  10 \alpha+10\gamma \\
		\hline
		(000010000110)& 12 \alpha+10\gamma  &(000010001001)  & 10\alpha-5\beta+10\gamma  \\
		\hline
		(000010000005)& 10 \alpha+10\beta+10\gamma  & (000010010000) & 10 \alpha-10\beta+10\gamma  \\
		\hline
	\end{array}$
\end{table}

\FloatBarrier

In the fifth power, again,  there is a number of diagrams, which  do not appear in the list of vertical sums of equal area $su$ diagrams, beside those which appear in powers of adjoint $\leq 4$ . First we list diagrams of area 10 below together with their value of Casimir, next in the same row, and next in  each row is some other diagram with the same Casimir, but already in the image of the map. It means that Casimirs of these out of the image diagrams are already covered by the universal Casimirs.

\begin{eqnarray}
	\ydiagram{4,4,2} \,\,  , C=10N, \,\, \ydiagram{5,3,1,1} \\
	\,\,\ydiagram{4,3,3} ,\,\,  C=10N-4, \,\,  \ydiagram{4,4,1,1} \\
	\ydiagram{4,2,2,2} ,\,\,  C=10N-16, \,\,  \ydiagram{4,3,1,1,1} \\
	\ydiagram{3,3,3,1},  \,\,  C=10N-16, \,\,  \ydiagram{4,3,1,1,1} \\
	\ydiagram{2,2,2,2,2} \,\,  C=10N-40, \,\,  \ydiagram{4,1,1,1,1,1,1}
\end{eqnarray}

Besides these five diagrams of area 10, there are following three diagrams of smaller area, which didn't appear in the previous powers of adjoint. We list them in the left column  below, next are their Casimirs, next are universal Casimirs, which give that values, next are $su$ representations (by Dynkin labels) giving that universal Casimir. Last two columns are not unique - one can find sometimes another universal Casimir with the same value on the $so$. Note however, that these universal Casimirs are those arising in the fifth power of adjoint, which is natural since diagrams in the first column arose in the fifth power of adjoint.

\begin{eqnarray}
	\ydiagram{5,3}, \quad C&=&8N+12, \quad  C=8\gamma+10\beta-2\alpha, (10010...01001) \\
	\ydiagram{5,2,1}, \quad C&=&8N+6, \quad C=8\gamma+10\beta+\alpha, (0110...01001)_s \\
	\ydiagram{5,1}, \quad C&=&6N+12, \quad C=6\gamma+10\beta+2\alpha, \quad (0110...0110)
\end{eqnarray}

Finally, there is one more diagram not in the image of the map $M$, whose Casimir however do not appear from the all listed universal Casimirs:

\begin{eqnarray}\label{5out}
	\ydiagram{5,1,1,1}, \quad C&=&8N, \quad 
\end{eqnarray}

This leads to a new universal Casimir, as we suggest in the next section. 

\subsection{Universal Casimirs eigenvalues for exceptional algebras}

We calculate (with the help of LieART package of Mathematica\texttrademark ) the spectra of Casimir eigenvalues on fifth power of adjoint of exceptional algebras, and compare it with the spectra of our universal Casimirs from table \ref{Cas5}. For $E_6$ all eigenvalues are contained in the universal eigenvalues, for $G_2, E_7$ absent are  eigenvalues of universal Casimirs on the  lower powers of adjoint, only. For $F_4$ and $E_8$ we get one eigenvalue for each, which is not covered by universal Casimirs and lower powers of adjoint. Namely, Casimir's eigenvalue 92 of representation {\bf 1 341 522} of algebra $F_4$, and 288 of representation {\bf 4 825 673 125} of algebra $E_8$. 

Both eigenvalues, as well as Casimir of out-of-image $so$ diagram (\ref{5out}) are covered by additional universal Casimir

\begin{eqnarray}
	C=10\gamma+8\beta +4\alpha
\end{eqnarray}
and its permutations. This universal Casimir satisfies condition to come from fifth power (due to coefficient 10 at $\gamma$), and remaining two coefficients have to satisfy at least four conditions: two for coincidence with Casimir eigenvalue for $so$ diagram (\ref{5out}), and two for $F_4$ and $E_8$ eigenvalues, so its existence was not guaranteed. 
We add this universal Casimir to the Casimirs of table \ref{tab:ad5} and suggest the result as a complete list of universal Casimirs, which can be used to get the complete universal decomposition of the fifth power of adjoint, along the lines of \cite{AIKM,IK25,IKP23}, see also  \cite{MS25,BMM25}.

\section{Conjectures and conclusion} \label{Conj}

The number of $so$ diagram at $k$-th power of adjoint is $p(2k)$, i.e. the number of partitions of $2k$. The number of $su$ diagrams in the same order is $p(k)(p(k)+1)/2$, since we can arbitrarily choose two diagrams of area $k$, and order is inessential.   At large $k$ $p(k)$ is of order

\begin{eqnarray}
	p(k) \sim e^{\pi\sqrt{2k/3}}
\end{eqnarray}

It follows that at large $k$ the number of $su$ diagrams is much larger than the number of $so$ diagrams: 

\begin{eqnarray}
	p(2k) \sim e^{2\pi\sqrt{k/3}} \\ \label{sugrow}
	p(k)(p(k)+1)/2 \sim e^{2\pi\sqrt{2k/3}}
\end{eqnarray}

So our map $M$ is sending the large number of $su$ diagrams into the small number of $so$ diagrams. However, we can always find $so$ diagrams which have no preimage, and the number of such diagrams obviously will be growing, also. Particularly, it is easy to find the number of diagrams in sixth power of adjoint, with area 12, which cannot be obtained as a vertical sum of diagrams with area 6 - it is equal to 5, the same number in the seventh power of adjoint is 13. Of course the number of these diagrams is much smaller than the exponentially growing number of 
$su/so$ diagrams. 

Correspondingly, it is reasonable to assume that  the number of additional universal Casimirs will grow, too, although not so fast as  the number of $su$ diagrams (\ref{sugrow}). We suggest the following 

 {\bf Conjecture 1.}

{\it All eigenvalues of Casimir operator on the given power of adjoint representation can be represented in universal form $C=x\alpha+y\beta+z\gamma$. The "most" part of these eigenvalues can be found by procedure described above, i.e. for each $su$ diagram one obtain corresponding $so$ diagram by the map $M$, then obtain the corresponding universal Casimir by (\ref{UCus}). Besides that there is a "small" number of universal Casimir eigenvalues,  arising from the $so$ diagrams out of the image of the map M. } 

Actually, we have one more conjecture to study. Our map $M$ is sending $su$ diagrams into those of $so$, this seems to have similarity with the folding procedure which folds Dynkin diagrams with symmetries (such as $A_n$) into other diagrams, e.g. $A_{2n-1} \rightarrow C_n$. Based on the lower area diagrams examples, we suggest 

 {\bf Conjecture 2.}
 
 {\it The vertical sum of Young diagrams is the (dual version of the) folding map of $A_n$.  }
 
 This conjecture explains   $Z_2$ symmetry of the universal formulae specialized to $su$ algebra, first observed in \cite{Cohen}, since the folding map is acting on the $Z_2$-invariant highest weights. More detailed version of this conjecture hopefully will explain other similar symmetries in the universal formulae, specialized to $D_n, n>4, E_6$ ($Z_2$ symmetry), and $D_4$ ($S_3$ symmetry) algebras. 

\section{Acknowledgments}

This work is partially supported by the Science Committee of the Ministry of Science  and Education of the Republic of Armenia under contracts   21AG-1C060 and 24WS-1C031.

\appendix

\section{Vogel's table}

Below is Vogel's table, listing points in projective plane, corresponding to simple Lie algebras. Homogeneous coordinates $\alpha, \beta, \gamma$ on projective plane are relevant up to the rescaling and permutations. 

\begin{table}[h] \caption{Vogel's  parameters for simple Lie algebras}     \label{tab:Vogel}
	\begin{tabular}{|r|r|r|r|r|r|} 
		\hline Algebra/Parameters & $\alpha$ &$\beta$  &$\gamma$  & $t=\alpha+\beta+\gamma$ & Line \\ 
		\hline $su(N)$ & -2 & 2 & $N$ & $N$ & $\alpha+\beta=0 $\\ 
		\hline $so(N), sp(-N)$ & -2  & 4 & $N-4$ & $N-2$ & $\alpha+2\beta=0$ \\ 
		\hline $Exc(n)$ & -2 & $n+4$  & $2n+4$ & $3n+6$& $\gamma=2(\alpha+\beta)$ \\ 
		\hline 
	\end{tabular} 
\end{table}

In Table \ref{tab:Vogel} for $su(N)$ and $so(N)$ $N$ is positive integer, for $sp(-N)$ $N$ is negative even integer, for exceptional line $Exc(n)$ $n=-1,-2/3,0,1,2,4,8$ for  $A_2,G_2, D_4, F_4, E_6, E_7, E_8$  respectively.

\end{document}